\pdfoutput=1
\documentclass[fleqn,10pt]{SelfArx}

\usepackage[T1]{fontenc}
\usepackage{textcomp}
\usepackage{array}
\usepackage{tabularx}
\usepackage{cite}
\usepackage{enumitem}

\makeatletter
\let\SA@thebibliography\thebibliography
\renewcommand{\thebibliography}[1]{%
  \SA@thebibliography{#1}%
  \footnotesize\sloppy
  \Urlmuskip=0mu plus 3mu\relax
  \def\UrlBreaks{\do\/\do-\do.\do\_\do\?\do\=}}
\makeatother

\definecolor{color1}{RGB}{0,0,90}
\definecolor{color2}{RGB}{0,20,20}

\usepackage{hyperref}
\usepackage{xurl}
\hypersetup{hidelinks,colorlinks,breaklinks=true,urlcolor=color2,citecolor=color1,linkcolor=color1,bookmarksopen=false,%
  pdftitle={A Multimodal Framework for Measuring Creator Communication on Short-Form Video},%
  pdfauthor={Ancis, Fire, Tarshish}}

\renewcommand{\thesection}{\Roman{section}}
\renewcommand{\thesubsection}{\thesection-\Alph{subsection}}

\titleformat{\section}
  {\color{color1}\large\sffamily\bfseries}
  {}
  {0em}
  {\colorbox{color2!10}{\parbox{\dimexpr\linewidth-2\fboxsep\relax}{\centering\thesection. #1}}}
  []

\newcommand{\sheading}[1]{%
  \phantomsection\addcontentsline{toc}{section}{#1}\section*{#1}}
\newcommand{\ssubheading}[1]{%
  \phantomsection\addcontentsline{toc}{subsection}{#1}\subsection*{#1}}

\newcounter{appx}
\renewcommand{\theappx}{\Alph{appx}}
\newcommand{\appheading}[2]{%
  \refstepcounter{appx}%
  \ssubheading{Appendix \theappx. #1}%
  \label{#2}}

\newenvironment{promptblock}
  {\begin{list}{}{%
     \setlength{\leftmargin}{1em}\setlength{\rightmargin}{0em}%
     \setlength{\listparindent}{0pt}\setlength{\itemindent}{0pt}%
     \setlength{\labelwidth}{0pt}\setlength{\labelsep}{0pt}%
     \setlength{\topsep}{0.5\baselineskip}\setlength{\parsep}{0.45\baselineskip}}%
   \item\relax\footnotesize\raggedright}
  {\end{list}}

\newlength{\pinunit}
\newenvironment{pseudo}
  {\par\addvspace{4pt}\noindent\begin{minipage}{\linewidth}\small\raggedright
   \setlength{\parindent}{0pt}\setlength{\parskip}{0pt}\setlength{\leftskip}{0.5em}%
   \everypar{\hangindent=2em\hangafter=1}}
  {\par\end{minipage}\par\addvspace{4pt}}
\newcommand{\pin}[1]{\hspace*{#1\pinunit}}

\JournalInfo{Preprint}
\Archive{}

\PaperTitle{A Multimodal Framework for Measuring Creator Communication on Short-Form Video}

\Authors{Julie R. Ancis\textsuperscript{1}, Michael Fire\textsuperscript{2}, Omer Tarshish\textsuperscript{2}}
\affiliation{\textsuperscript{1}\textit{Department of Informatics, New Jersey Institute of Technology, Newark, NJ, USA}}
\affiliation{\textsuperscript{2}\textit{Department of Computer Science, Ben-Gurion University of the Negev, Beer-Sheva, Israel}}

\newcommand{\keywordname}{Keywords}

\Abstract{
Short-form video combines speech, imagery, and on-screen text, creating challenges for systematic content analysis. We present a five-stage multimodal framework for measuring creators' themes and communication strategies at scale. The framework integrates platform metadata, speech transcripts, and vision--language descriptions of sampled frames, then applies large language models to generate predefined indicators and open-ended labels. An audited keyword scheme consolidates thematic labels, while creator-weighted aggregation gives each account equal weight regardless of posting volume. We demonstrate the framework on 9,852 TikTok videos from sixteen creators, retaining 7,311 videos for analysis across seven themes and seven communication indicators. Automated profiles reveal substantial differences in creators' use of education, humor, and responses to current events. Validation against two independent expert annotators yielded a mean framework--expert agreement of Cohen's $\kappa = 0.40$ across the six binary indicators, close to the $\kappa = 0.42$ the two experts achieved with each other, though agreement varied substantially across individual indicators. The framework provides a modular approach to comparative analysis of creator communication, combining scalable annotation with transparent aggregation and construct-specific validation.
}

\begin{document}

\flushbottom

\maketitle

%\tableofcontents

\thispagestyle{empty}

\section{Introduction}
\label{sec:intro}

Social media influencers now shape what large audiences see and believe. They build followings through reciprocity and intimacy, and the credibility that follows transfers to the products, values, and ideas they endorse \cite{maropo2020ifelt,reinikainen2020great}. Their influence extends well beyond consumption to sustainable habits and advocacy \cite{confetto2023sustainability}, political participation \cite{valenzuela2013unpacking}, and health behavior \cite{aljunaid2024unpacking,wasike2022influencer}. Creators also mobilize followers for collective action in movements such as Black Lives Matter and \#StopAsianHate \cite{shahin2025stopasianhate}. Much of this now happens in short videos, on platforms such as TikTok, whose feed carries a large share of Generation Z's popular culture \cite{cervi2021tiktok}.

Measuring what creators produce on social media is a content-analysis problem that does not scale. Manual coding is reliable but costly, so studies of creator practice tend to stay relatively small and interpretive. At the same time, the computational alternatives are built for detection: a classifier predicts a single target label, such as whether a post carries misinformation or harmful content \cite{pramanick2021momenta}, and says nothing about the repertoire of topics and strategies a creator actually uses. Short-form video compounds the difficulty because meaning is carried simultaneously by speech, on-screen text, imagery, and platform convention. Hence, a caption or a transcript alone recovers only part of the message.

Studies that measure a video across several dimensions at once already exist, but such breadth has rarely been paired with a large corpus. Manual content analysis codes every video against a checklist and reports prevalence per feature, as in a study of 100 trending videos from a COVID-19 vaccine hashtag \cite{basch2021global}. Automated designs reach more videos, and political TikTok has been characterized using computer vision, natural language processing, and statistical tools \cite{medinaserrano2020dancing}. Still, each construct in such designs rests on a tool trained for it, so the number of measures remains small whatever the corpus size.

Multimodal large language models change that. A single model can process sampled frames, a transcript, and a written instruction together and answer in free text \cite{bai2025qwenvl}. A new indicator can be added simply by describing it, without collecting training data for it, and the model can be asked what a video is about without being restricted to a fixed set of predefined answers. Because no human has to code each video, the full set of questions can be applied to every video in an archive rather than to a hand-coded sample. Applications to social media are recent and still few in number. One study used a multimodal model to grasp four abstract concepts across 142 depression-related YouTube Shorts, comparing its interpretations with those of human coders~\cite{liu2025can}. Another used vision--language models to characterize the videos served by a TikTok recommender \cite{masood2026counting}. Using such models to measure such indicators across a creator population remains unexplored.

In this study, we present a multimodal computational framework that converts short-form videos published by creators on social media platforms into comparable creator-level measures of theme and communicative strategy. Given a set of target accounts, the framework executes five sequential stages: (1) \textit{Data collection} retrieves each account's public video archive and platform metadata; (2) \textit{Multimodal enrichment} renders every video into a single text document joining a metadata header, an automatic speech transcript, and prompted vision--language descriptions of sampled frames; (3) \textit{Content classification} submits that document to a large language model (LLM) twice, once against a set of indicators fixed in advance and once with open-ended labels that allow themes to emerge inductively; (4) \textit{Taxonomy construction} resolves the resulting vocabulary onto master categories through a keyword scheme refined in an audited loop; and (5) \textit{Normalization and aggregation} turns the per-video codings into a prevalence for each creator, then averages across creators so that the busiest accounts do not dominate the result.

The architecture is independent of what is being measured. Three things are supplied per study: the target account list, the fixed indicator questions and response scales, and the master categories with their keyword definitions. The framework is modular by construction, since each stage is defined by the product it passes to the next. To test the framework, we applied it at scale to the work of fighting online hate speech, where antisemitism is among the most persistent forms \cite{ancis2025cyberpsychological}. There, narrative intervention by influencers, who combine reach with credibility, is a promising response  \cite{ozalp2020antisemitism}. The measurement task is to characterize what a population of creators does. Which themes do these creators address, and what is the prevalence of those themes? Which communicative and platform strategies do they use, and where do creators converge or diverge? And, how far can language-model coding be trusted to answer such questions?

We demonstrate the framework on the video archives of sixteen creators. Educational content is the most common communication strategy, measured by closed indicators, ahead of responses to current events, humor, direct calls to action, and personal storytelling. Two kinds of creators recur among the sixteen. One kind teaches, combining educational content with close tracking of the news cycle. The other entertains, building the same advocacy on humor, and one of these creators uses humor in 99\% of their videos. Two domain experts with doctoral-level training in psychology coded a sample of the corpus, blind to all model output. For the closed questions, the automated coding agrees with the experts about as strongly as the experts agree with each other. Averaged over the two annotators, chance-corrected agreement (Cohen's $\kappa$) is 0.40, against 0.42 between the annotators themselves, a level that reflects how hard these judgments are for people as well. For four of the six closed indicators, the coding more closely agrees with one expert than the experts agree with one another. Agreement is strongest where the evidence is plain in a single video. Against one expert, the humor-or-sarcasm indicator reaches 0.70, compared to the 0.71 that the experts reach with each other. Against the other expert, the October 7 thematic category reaches 0.75, with agreement on 92\% of videos.

This study makes four key contributions:
\begin{itemize}[leftmargin=1.2em, itemsep=2pt, topsep=3pt, parsep=0pt]

\item \textbf{We introduce a multimodal framework for measuring creator communication.} We integrate platform metadata, speech transcripts, and visual descriptions with predefined indicators and open-ended thematic coding. Creator-weighted aggregation enables systematic comparisons without allowing prolific accounts to dominate the results.

\item \textbf{We release a reusable open implementation.} Our pipeline, which will be publicly available upon publication, supports end-to-end analysis of creator video archives. Its modular design allows researchers to replace individual models and adapt the account list, indicator questions, and category scheme to new research settings.

\item \textbf{We characterize influencer communication at scale.} We analyze 9,852 TikTok videos from sixteen creators active in Jewish advocacy and countering antisemitism. Across the 7,311 retained videos, we quantify seven themes, six strategy indicators, and emotional intensity, revealing substantial variation in creators' topics and strategies.

\item \textbf{We benchmark automated coding against expert judgments.} We compare model outputs with independent, blinded annotations from two domain experts. For the six binary indicators, mean model--expert agreement is $\kappa = 0.40$, compared with $\kappa = 0.42$ between experts. Per-indicator comparisons identify which measures align most closely with expert coding.

\end{itemize}

The remainder of the paper is organized as follows. Section~\ref{sec:related} reviews the two literatures on which the framework is based, computational analysis of multimodal social media content and influencer narratives. Section~\ref{sec:methods} specifies the five stages. Section~\ref{sec:experiments} describes the study corpus and the configuration of each stage. Section~\ref{sec:results} reports the creator-level measurements. Section~\ref{sec:discussion} discusses the findings and limitations, and Section~\ref{sec:conclusions} concludes with directions for future work.

\section{Related Works}
\label{sec:related}

\subsection{Computational analysis of multimodal social media content}

Multimodal content analysis, which examines how text, images, video, and audio combine to create meaning in social media posts, captures richer signals compared to text-only approaches. This analysis allows for an understanding of how different modes interact by reinforcing, contradicting, or adding nuance to each other. Combining visual and textual cues consistently improves accuracy over text-alone methods \cite{gandhi2023multimodal,niu2016sentiment,vempala2019categorizing}. Multimodal models have been applied to misinformation and fake news detection \cite{nakamura2020fakeddit,zlatkova2019factchecking,zhou2020safe} and to detecting harmful content \cite{pramanick2021momenta}.

These multimodal classifiers for sentiment analysis, misinformation, fake news, and harmful content share a task shape: each predicts one target label per item. Measuring what a creator does is a different problem. It requires characterizing many descriptive constructs simultaneously and comparing them across a population of creators.

A more closely related line of work uses large language models as coders. Zero-shot annotation from natural language instructions has been shown to match or exceed crowd workers on several text labeling tasks \cite{gilardi2023chatgpt} and, on political social media messages, to outperform expert coders and supervised classifiers as well \cite{tornberg2025llm}. A systematic evaluation across twenty-five computational social science benchmarks refines this picture: zero-shot LLMs fall short of fine-tuned classifiers on taxonomic labeling, yet their free-form output can exceed  crowd-worker references in quality \cite{ziems2024can}. The present design builds on this distinction, pairing a closed pass for comparability, at some cost in accuracy, with an open pass for inductive description.

\subsection{Influencer narratives: the demonstration case}

Social media influencers (SMIs) are a powerful yet underexamined intervention against online hate. Creators who promote brands and lifestyles \cite{lajnef2023effect} measurably shift attitudes, knowledge, and behavior, from sustainable habits and advocacy \cite{confetto2023sustainability} to political action \cite{valenzuela2013unpacking}, for good and ill alike \cite{aljunaid2024unpacking,wasike2022influencer}. The mechanism is parasocial: repeated, seemingly personal interaction builds credibility, trust, and loyalty which transfer to the values and ideas a creator endorses \cite{maropo2020ifelt,reinikainen2020great}.

That capacity extends into the direct action of followers. Influencers mobilized followers for the Black Lives Matter movement with personal stories, educational content, and calls to action, and \#StopAsianHate gained traction partly through prominent Asian celebrities, whose participation amplified the movement even as right-wing accounts tried to hijack the hashtag \cite{shahin2025stopasianhate}. The dynamic is sharpest on TikTok, whose participatory, algorithmically amplified feed reflects the popular culture of Generation Z \cite{cervi2021tiktok}. It is no more immune to hate speech than any other platform, and some creators use their reach to advocate for the groups such hate speech targets.

One structured form of that work is counterspeech: responses to rhetoric that is clearly harmful but may fall below the legal threshold for restriction or removal \cite{cypris2024effectiveness}. Rather than silencing speech, counterspeech challenges or delegitimizes harmful narratives through factual correction, exposure of hypocrisy or contradiction, warning of consequences, emotional appeals, denunciation, and humor or sarcasm \cite{benesch2016considerations}. 

Hate speech is a measurable presence across languages and national contexts, though a relatively small share of total content \cite{tonneau2025hateday,siegel2021trumping}, and antisemitism occupies a prominent and growing position within it~\cite{ancis2025cyberpsychological}: roughly 9--27\% of flagged comments in large UK, French, and German corpora, rising sharply after the October 7, 2023 Hamas attacks \cite{becker2023celebrating,becker2024decoding}; a 360\% surge in US incidents after those attacks \cite{adl2024incidents}; more than 150,000 antisemitic posts from over 1,000 US-based extremist accounts in six months, led by conspiratorial content \cite{isd2025antisemitism}; and 73\% of American Jews reporting encounters with antisemitism online \cite{ajc2025state}. The open question is not whether it propagates but how it might be countered, including by influencers who leverage reach and credibility to challenge hate \cite{ancis2025countering}.

Hate speech inflicts documented harm on its targets, whose reactions follow the stages seen in other traumatic experiences \cite{leets2002experiencing}. The promise and limits of counterspeech as a response are debated at length in the conceptual literature \cite{cepollaro2023counterspeech,lepoutre2017hate}. Effects on bystanders and perpetrators have also been studied \cite{cypris2024effectiveness,jia2025tackling}. Among bystanders, participants who saw counterspeech from other users were more likely to speak up than those who saw only neutral replies, suggesting that counterspeech can shift perceived norms and encourage collective action. Effectiveness with perpetrators depends partly on the speaker: those with larger followings, or who  belong to the perpetrator's in-group, are more likely to change behavior \cite{chung2024understanding}.

Similarly, counter-narratives to antisemitic content on Twitter that were posted by credible actors, such as community organizations, spread further and lasted longer than the antagonistic content they addressed, a pattern the authors interpret as collective efficacy \cite{ozalp2020antisemitism}. SMIs combine precisely the properties these findings point to: audience reach, perceived credibility, and salient identity cues.

The research on influencers and the research on counterspeech both bear directly on the framework reported here. The finding that a speaker's credibility and visibility shape how far narratives travel \cite{ozalp2020antisemitism,chung2024understanding} is why prevalence is reported per creator rather than pooled over videos: the unit of interest is the practice of a creator, not the volume of an archive. The counterspeech repertoire cataloged by Benesch \emph{et al.} \cite{benesch2016considerations} is the kind of construct a fixed question set can capture. The set used here covers part of that repertoire (humor or sarcasm, appeals to data, and emotional intensity) and leaves other elements, such as exposure of hypocrisy and warnings of consequences, outside its scope. From content analysis, the framework inherits a two-pass design: a closed pass, in which questions are fixed in advance to support comparison across creators, and an open pass, in which themes emerge from the material rather than from the codebook. Finally, the evidence that visual and textual cues together outperform text alone \cite{gandhi2023multimodal,niu2016sentiment} justifies building one document per video from a transcript and frame descriptions rather than relying on captions alone.

\section{Methods}
\label{sec:methods}

This study presents a creator-agnostic, multimodal computational framework for the systematic analysis of short-form video content on TikTok. Taking a finite set of target influencer accounts as input, the framework produces free-form thematic labels and structured rhetorical and strategic indicators that support analysis of both individual creators and the corpus as a whole. Throughout this section, we illustrate each stage with a single running example: a 19-second video published in April 2023.\footnote{Published by @hen.mazzig on April 3, 2023 (\url{https://www.tiktok.com/@hen.mazzig/video/7217884208163785986})}

\subsection{Framework Overview}

\begin{figure*}[!t]
  \centering
  \includegraphics[width=\textwidth]{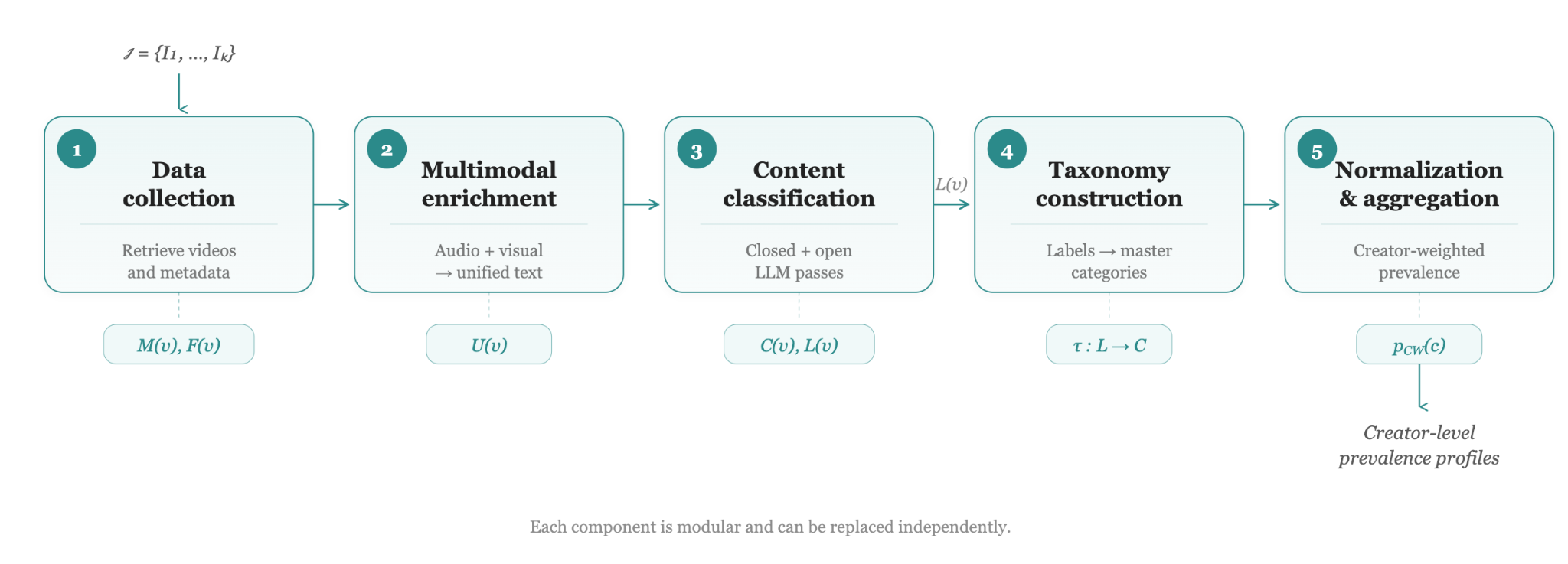}
  \caption{The five stages of the framework. Each stage passes a named product to the next: the archive and metadata $M(v), F(v)$, the unified document $U(v)$, the closed and open codings $C(v), L(v)$, the frozen label-to-category map $\tau : L \to C$, and the creator-weighted prevalence $p_{\mathrm{CW}}(c)$. The three models occupy replaceable slots.}
  \label{fig:framework}
\end{figure*}

Let $\mathcal{I} = \{I_1, \ldots, I_K\}$ denote the set of target influencer accounts, and let $V_k$ denote the set of public videos posted by influencer $I_k$. For each video $v \in V_k$, the framework executes five sequential stages (see Figure~\ref{fig:framework}): (1) data collection, (2) multimodal enrichment, (3) content classification, (4) taxonomy construction, and (5) normalization and aggregation. The framework is modular: the speech-to-text model, the vision--language model, and the LLM used for classification can each be replaced without altering the analytical logic.

Stage 1 - \textbf{Data collection}. Given an influencer $I_k$, the full set of publicly accessible videos $V_k$ is retrieved via web crawling, platform API, or a third-party collection tool. For each video $v \in V_k$, two data items are collected: the platform metadata $M(v)$ and the raw video file $F(v)$. The metadata $M(v)$ comprises the fields exposed by the platform at crawl time, such as the publication timestamp, caption, hashtags, and engagement counts (e.g., play count, likes, comments, shares, and saves). The exact set of fields retrieved in our study is specified in Section~\ref{sec:experiments}. The raw file $F(v)$ is downloaded once and serves as the source for both the audio and visual modalities processed in Stage 2. To ensure dataset integrity, reposts and duplicates are filtered during collection, with videos deduplicated using their unique platform identifiers. This yields a clean, deduplicated archive $\{(M(v), F(v)) : v \in V_k\}$ for each influencer $k = 1, \ldots, K$. For the running example, the metadata record $M(v)$ captures the publication timestamp (2023-04-03 17:57:15 UTC), a short caption carrying seven hashtags (``Happy \#Passover!\ldots''), and the engagement counts at crawl time (1,111 plays, 98 likes, 10 comments, 5 shares, and 8 saves), while the raw file $F(v)$ is a 19.07-second clip at $576 \times 1{,}024$ pixels and 30 fps.

Stage 2 - \textbf{Multimodal enrichment}. This stage transforms each video \emph{v} into a single unified textual document $U(v)$ suitable for LLM-based classification. The audio track is extracted from $F(v)$ and passed to a speech-to-text model, producing a transcript $T(v)$. Outputs are post-processed to remove known speech-to-text artifacts, such as a ``Thank you for watching'' tag appended by the model when no such phrase was spoken. In parallel, frames are sampled at fixed intervals along the video timeline, and each sampled frame is passed to a vision--language model with a task-specific prompt that directs it to describe the people, on-screen text, setting, mood, and any brand or cultural elements present in the frame. The descriptions are not default model captions but are elicited by this structured prompt, which is reproduced in Appendix~\ref{app:vlprompt}. To keep descriptions of the same video internally consistent, a short carry-forward summary of person, brand, and setting features recovered from earlier frames is appended to the prompt for each subsequent frame, so that frames of a single video are described in sequence rather than in full mutual independence. The per-frame descriptions are concatenated in chronological order, each tagged with its approximate timestamp, to form the visual-description component $D(v)$. The unified document is then assembled as $U(v) = \mathrm{HEADER}(M(v)) \oplus T(v) \oplus D(v)$, where $\oplus$ denotes textual concatenation. The header lists the creator handle, video identifier, publication date, caption, hashtags, and engagement metrics; the transcript follows in continuous prose, and the visual descriptions follow in chronological sequence. For the running example, the speech-to-text model produces a 56-word English transcript that opens ``Our ancestors ended centuries of slavery and split the sea in half just so we could be free today,'' and twenty frames sampled at 1-second intervals from $t = 0$ to $t = 19$, each yield a structured natural-language description. For instance, the frame at $t = 17$ is described as showing ``a white t-shirt with a necklace featuring a Star of David pendant,'' with ``the words `HAPPY PASSOVER' written in colorful, playful letters'' on the shirt, visual details absent from the transcript. The resulting $U(v)$ comprises the metadata header, this transcript, and the twenty frame descriptions.

Stage 3 - \textbf{Content classification}. This stage operates on $U(v)$ and applies an LLM with a flexible configuration that can be adjusted as needed, together with a structured response format. The full prompt templates and response schemas are provided in Appendix~\ref{app:prompts}. Two complementary passes are executed over the same unified document.

The closed-question pass evaluates each video against a fixed set of \emph{m} content-strategy indicators defined a priori based on the relevant literature. For instance, one indicator evaluates whether the creator employs personal storytelling, another whether the video uses humor or sarcasm, and a third whether it references current events. For each indicator, the model returns a categorical response, for example, Yes, No, or Unclear. Presented with $U(v)$ and a fixed instruction template, the model produces a structured response $C(v) = \{c_1, \ldots, c_m\}$, where \emph{m} is the fixed number of closed questions and each component takes a value from a predefined response scale. This pass supports systematic cross-creator comparison along fixed dimensions. For the running example, the closed pass returns, among others, $c_2 =$ ``Yes'' for educational content and $c_{6} =$ ``Yes'' for strategic hashtag use, while $c_1$ for personal storytelling and $c_5$ for humor or sarcasm both return ``No.''

The open-question pass yields $L(v) = \{l_1, \ldots, l_{n(v)}\}$, a set of short, free-form labels describing the strategies and themes of the video. The decision to leave this pass open-ended rather than constraining it to a predefined set of labels serves two purposes: it avoids biasing the model toward categories that may not adequately capture the content, and it aligns with established content-analysis methodology in which coders are asked open-ended questions to allow themes to emerge inductively from the data. Here, the model is presented with $U(v)$ and a uniform extraction prompt with a per-label length constraint that keeps labels discrete and analytically tractable. For the running example, the open pass returns $n(v) = 11$ labels, including ``Passover holiday greeting'' and ``Exodus slavery narrative,'' which describe its topics, and ``Motivational religious message'' and ``Direct-to-camera monologue,'' which describe its style and strategy. Taken together, the two passes are complementary. The closed pass provides comparability along fixed dimensions, while the open pass provides thematic depth and inductive discovery.

Stage 4 - \textbf{Taxonomy construction.} The open-question pass of Stage 3 returns, for each video, a set of free-text labels $L(v)$. Because the pass is deliberately open-ended, the labels form a large, uncontrolled vocabulary in which a single recurring practice appears under many near-duplicate phrasings. For example, ``direct-to-camera monologue,'' ``direct address to camera,'' and ``direct-to-camera delivery'' all refer to the same practice. In this raw form, the labels cannot be counted or compared across videos. This stage resolves the vocabulary onto a fixed, interpretable set of \emph{M} master categories, serving as the bridge between the inductive output of classification and the quantitative analysis that follows.

The mapping is constructed once, at the corpus level, in four main steps, and the frozen result is then applied unchanged to every video. First, the labels of all videos are pooled into a single vocabulary and ranked by the number of videos in which each label occurs, which concentrates attention on the small set of high-frequency labels that account for most of the corpus mass. Second, a provisional set of master categories is specified from two complementary sources. Some categories are derived deductively from the study's aims and prior literature \cite{ancis2025countering}. Others are created inductively, by promoting recurrent, high-frequency labels that fall outside the initial structure into categories of their own. The resulting scheme is theoretically informed yet grounded in the vocabulary actually observed. Third, each category is operationalized as a set of keyword stems, base word forms such as ``humor'' or ``lifestyle,'' and a label is assigned to a category whenever one of the category's stems occurs in it, matched case-insensitively so that a stem such as ``educat'' captures ``education,'' ``educational,'' and ``educating'' alike. Fourth, the stem sets are refined in an iterative audit loop. Each pass corrects two error types: stems are added where labels are systematically missed, and overly broad stems are constrained to whole-word matches where they capture unrelated terms. When a full pass produces no further change, the category definitions are frozen. Because every assignment records the stem that triggered it, each category membership is traceable to its lexical cause. The formal definitions, the refinement pseudocode, and the error types corrected at each pass are given in Appendix~\ref{app:taxonomy}.

Assignment is non-exclusive. A label may match several categories, and a video is assigned to every category matched by any of its labels, $\mathrm{cat}(v) = \{C_m :$ there exists $l \in L(v)$ which is assigned to $C_m\}$. This per-video category membership, rather than the raw label text, is what passes to Stage 5. For the running example, ``Passover holiday greeting'' matches the stem ``passover'' and is assigned to Jewish Identity / Culture / Pride, and ``Jewish historical memory'' matches both ``jewish'' and ``historical memory,'' so it is assigned to Jewish Identity / Culture / Pride and to Holocaust / Historical Memory. The video's other nine labels, such as the style label ``Direct-to-camera monologue,'' match none of the seven categories defined for this study (Table~\ref{tab:categories}) and are not mapped. The video therefore enters Stage 5 as an instance of these two categories, with its individual phrasings collapsed onto the shared scheme.

The categories are divided into substantive thematic domains and out-of-domain material. The relevance restriction retains every video whose labels map to at least one substantive category and excludes videos whose labels are confined to out-of-domain material, typically lifestyle or entertainment content. The retained set defines the analytic corpus on which all subsequent statistics are computed. The complete set of categories and their keyword definitions for our study is reported in Table~\ref{tab:categories}.

Stage 5 - \textbf{Normalization of creator-level category prevalence.} This stage takes the per-video category assignments from Stage 4 and produces the summary statistics used in the analysis. Two structural features of the data prevent naive aggregation of the raw assignments. The first is label redundancy: Stage 3 frequently produces multiple labels for a single video that all map to the same master category. The second is volume imbalance: creators vary widely in how many videos they post, so pooled statistics would reflect output volume rather than creators' practice. The normalization resolves these issues in three steps. First, redundancy is removed by reducing each video to binary presence per category. For each video \emph{v} and category \emph{c}, the indicator $I(v, c)$ equals 1 if any label of $v$ maps to $c$ and 0 otherwise. This follows standard content-analysis practice in which the unit of analysis either exhibits a code or does not. A video that yields several labels mapping to the same category is therefore counted once. Second, each creator's output is normalized into a prevalence profile. For creator \emph{k} with video set $V_k$, the proportion $p(k, c) = \frac{1}{|V_k|} \sum_{v \in V_k} I(v, c)$ is the share of creator $k$'s videos in which category $c$ is present. The profile is bounded on $[0, 1]$ and is invariant to the absolute number of videos, so all creators are placed on a common scale. Third, the creator profiles are combined using creator-weighted aggregation, which is adopted as the primary corpus-level statistic. The creator-weighted prevalence is $p_{\mathrm{CW}}(c) = \frac{1}{K} \sum_{k=1}^{K} p(k, c)$. In words, $p_{\mathrm{CW}}(c)$ is the equally weighted mean  of the $K$ creator-level proportions, so each creator contributes equally regardless of video volume. The three steps are summarized in pseudocode in Appendix~\ref{app:normalization}.

For each category, the creator-weighted prevalence $p_{\mathrm{CW}}(c)$ serves as the primary statistic. The standard deviation of $p(k, c)$ across the $K$ creators accompanies it as a measure of between-creator dispersion, showing where creators share a practice and where they diverge. The video-weighted pooled rate $p_{\mathrm{VW}}(c) = \frac{1}{N} \sum_{v \in V} I(v, c)$, where $V$ is the analytic corpus and $N = |V|$, is reported alongside as a robustness check. Because $p_{\mathrm{VW}}(c)$ weights every video equally while $p_{\mathrm{CW}}(c)$ weights every creator equally, the gap between the two estimates measures the effect of the normalization itself. The creator-weighted mean is an average of $K$ proportions, so each creator contributes exactly $p(k, c) / K$ to it. The results use this decomposition to show which creators drive each category.

The same normalization is applied to the closed-question indicators of Stage 3. Each categorical indicator is reduced to binary presence per video under a response rule specified by the study, and Steps 1 through 3 then produce per-creator proportions and creator-weighted means directly comparable to the thematic statistics. An ordinal indicator enters the same computation with its ordinal score in place of the binary indicator, so $p(k, c)$ becomes the creator's mean score and $p_{\mathrm{CW}}(c)$ its creator-weighted mean. Corpus-level descriptive statistics, including engagement distributions, posting timelines, and label-vocabulary characteristics, are reported alongside these normalized indicators in Section~\ref{sec:results}.

\textbf{Validation design.} The framework is validated against human coding of a sample drawn from the analytic corpus. Annotators work independently, blind to the model's output and to one another's codes. Model--annotator agreement is reported as raw percent agreement and Cohen's $\kappa$ on the binary coding used in the analysis. Agreement between annotators serves as a reference point for the difficulty of each judgment. For the ordinal emotional-intensity item, two further measures are reported: the share of videos on which the model's score and the annotator's fall within one point of each other, and Spearman's rank correlation between the two. The correlation is computed both across individual videos and across per-creator mean scores, since the creator is the unit of analysis.

\section{Experiments}
\label{sec:experiments}

This section reports a demonstration study in which the framework is applied end to end to a single corpus. Sixteen TikTok accounts active in Jewish advocacy and counter-antisemitism content were selected in two stages, their complete public archives were collected on June 4, 2025, and each stage was then instantiated with a specific model and configuration. The output was validated in two ways: against two expert annotators coding a stratified sample of 100 videos, and against three reduced versions of the same documents that isolate what the multimodal enrichment contributes. What follows sets out the account selection, the collected corpus, the instantiation of each stage, the aggregation rule, and the two validation procedures in turn. Throughout, the term `theme' is reserved for the master categories produced by the open pass of Stage 3 and resolved in Stage 4, and the term `indicator' is used only for the closed items.

\textbf{Account selection.} The sixteen accounts are the outcome of a two-stage selection process. In the first stage, the research team compiled a list of 23 candidate creators active in Jewish advocacy or counter-antisemitism content, identified from public advocacy activity on the platform. Candidates were chosen purposively for diversity on content domain, advocacy style, and creator background, so that no single genre, mode of address, or community would dominate the list. In the second stage, candidates were screened for an active TikTok presence with a publicly accessible, retrievable video archive at the time of collection. Seven candidates failed this criterion, leaving $K = 16$ accounts. The retained accounts preserve the intended diversity: their content ranges from  comedy, lifestyle, and fitness to targeted political commentary and religious experiences, and the accounts represent a broad demographic spectrum of voices, spanning Jewish, Muslim, and Christian creators, including Arab citizens of Israel. Based on TikTok follower counts gathered in October 2026, audiences range from roughly 19,000 followers (Hen Mazzig) to roughly 9.1 million (Montana Tucker). Using standard industry tiers, three of the creators have more than one million TikTok followers (mega), six have between 100,000 and one million (macro), and seven have fewer than 100,000 (micro). Several creators, including Hen Mazzig, Emily Schrader, and Lizzy Savetsky, have substantially larger followings on Instagram or X than on TikTok; however, this study focuses on TikTok videos, and accounts were classified by their TikTok following alone. TikTok was chosen because short-form video is its native and primary format, so each post combines several modes of communication at once: moving image, speech, on-screen text, written captions, music and sound, visual effects, and hashtags \cite{kaye2022tiktok}. Platform features such as duets, stitches, and reusable sounds also allow creators to build directly on one another's videos, adding a layer of meaning that text- or image-based posts on other platforms do not carry in the same way \cite{zulli2022extending}. In addition, because TikTok's For You feed distributes videos to users who do not follow the creator, follower count is an imperfect measure of a video's potential reach, and smaller accounts can reach audiences well beyond their own following. Including creators across these tiers and platform profiles allows the analysis to consider both high-reach accounts and smaller accounts with more concentrated audiences, rather than reflecting only the most visible voices. For each account, the full public video archive available at collection time, June 4, 2025, was retrieved using a third-party collection tool, along with the metadata required by Stage 1: the platform identifier, publication timestamp, caption, hashtags, engagement counters, duration, and audio track.

\textbf{Collected corpus.} The Stage 1 collection yielded 9,852 videos posted between April 3, 2018 and June 4, 2025 with per-creator volumes varying by more than an order of magnitude (Table~\ref{tab:collection}). Posting volume grows over this collection window: 3,277 videos, a third of the archive, were published in 2024 alone, the last full year before collection. The archive comprises 155 hours of footage, with a median video length of 43 seconds. Engagement follows the heavy-tailed distribution typical of platform data. The median video draws 28,400 plays while the mean is 270,342, a gap produced by a small set of viral videos. Median engagement rates vary across the sixteen accounts (Table~\ref{tab:collection}), from 3.4\% to 12.0\%. Engagement is measured per view, rather than by follower tier, because per view engagement is the measure the platform itself foregrounds.

\begin{table}[!t]
  \scriptsize
  \setlength{\tabcolsep}{3pt}
  \caption{Per-Creator Collection Summary.}
  \label{tab:collection}
  \centering
  \begin{tabular}{@{}lrlrrrr@{}}
    \toprule
    \textbf{Creator} & \textbf{Videos} & \textbf{First post} & \textbf{Retained} & \textbf{Retention} & \textbf{Med. plays} & \textbf{ER} \\
    \midrule
    yosephhaddad     & 2,325 & 2021-06 & 2,309 & 99.3\%  & 39,700  & 6.1\%  \\
    chloehechterr    & 1,453 & 2020-03 & 929   & 63.9\%  & 18,000  & 4.6\%  \\
    montanatucker    & 1,216 & 2018-04 & 283   & 23.3\%  & 875,750 & 9.2\%  \\
    shaharcohen9      & 775   & 2020-08 & 464   & 59.9\%  & 53,400  & 5.3\%  \\
    lizzysavetsky    & 727   & 2019-11 & 666   & 91.6\%  & 9,238   & 6.8\%  \\
    motiankari     & 603   & 2019-12 & 174   & 28.9\%  & 9,668   & 4.8\%  \\
    justhallel       & 422   & 2019-11 & 399   & 94.5\%  & 1,361   & 3.4\%  \\
    noatishby        & 405   & 2022-02 & 404   & 99.8\%  & 46,800  & 6.6\%  \\
    eitanthegoalie   & 400   & 2019-01 & 370   & 92.5\%  & 21,850  & 11.0\% \\
    zach.sage        & 367   & 2022-02 & 202   & 55.0\%  & 10,600  & 6.5\%  \\
    emilyintelaviv   & 266   & 2021-06 & 260   & 97.7\%  & 3,751   & 7.3\%  \\
    alexandramoulavi & 233   & 2020-06 & 221   & 94.8\%  & 24,400  & 6.5\%  \\
    rawanosman2024       & 221   & 2023-10 & 221   & 100.0\% & 39,600  & 4.9\%  \\
    hen.mazzig       & 201   & 2021-03 & 201   & 100.0\% & 1,984   & 7.8\%  \\
    rudyrochman    & 128   & 2020-05 & 127   & 99.2\%  & 22,900  & 6.9\%  \\
    natebuzz         & 110   & 2021-03 & 81    & 73.6\%  & 191,350 & 12.0\% \\
    \midrule
    \textbf{Total} & \textbf{9,852} & & \textbf{7,311} & \textbf{74.2\%} & \textbf{28,400} & \textbf{6.3\%}\\
    \bottomrule
  \end{tabular}
  \par\vspace{2pt}{\scriptsize Creators are listed by their TikTok handle on the collection date. The first post is the earliest video in the archive; every archive runs to the collection date of June 4, 2025. Retained is the number of videos passing the relevance restriction of Stage 4, and Retention is its share of Videos. ER is the median view-based engagement rate, computed as $\frac{likes + comments + shares}{views}.$}
\end{table}

\textbf{Pipeline instantiation.} Audio tracks were transcribed directly from the video files with Whisper large-v3 \cite{radford2023whisper}. Frames were sampled at one-second intervals, capped at 200 frames per video, resized to $448 \times 448$ pixels, and described by Qwen2.5-VL 7B \cite{bai2025qwenvl}. The transcript, frame descriptions, and metadata were assembled into the unified document $U(v)$ of Stage 2. Comment threads were excluded from the document so that classification reflects the creator's own content rather than audience reactions. To bound document length, frame descriptions in documents exceeding 20,000 characters were evenly subsampled to at most 30 frames. Classification was performed by GPT-5.1~\cite{openai2025gpt51} with structured JSON output under a strict schema and a 4,096-token completion limit. The two passes of Stage 3 were sent together, as two tasks in a single request per video, over the same unified document. The open-question pass produced the free-form label list for each video. The closed-question pass comprised eighteen questions, of which seven are reported here. The remaining eleven were answered Yes on fewer than 20\% of videos, eight of them on fewer than 10\%, yielding too few positive cases to support comparisons across creators. The seven retained items comprise six binary indicators, with Yes rates from 33.6\% to 60.1\%, and one ordinal item, emotional intensity. The response scales are presented in Table~\ref{tab:indicators} and the full wordings in Appendix~\ref{app:prompts}.

\textbf{Aggregation.} The normalization of Stage 5 was instantiated with $K = 16$, the number of accounts in the corpus. Because some behaviors are only weakly observable in a single posted video, a conservative rule was applied throughout. A video counts as exhibiting an indicator only when the model answered Yes; both No and Unclear are treated as absence. Every reported indicator prevalence is therefore conservative with respect to undecided cases, since videos the model could not classify are counted as not exhibiting the behavior. Emotional intensity is an ordinal item that admits no Unclear response, so each creator's value is the mean of the scores across their videos. The three quantities defined in Stage 5 appear in Tables~\ref{tab:theme-prevalence} and~\ref{tab:indicators} as CW (creator-weighted mean prevalence), SD (standard deviation across creators), and VW (video-weighted pooled rate).

\textbf{Human validation.} A random sample of 100 videos, stratified by creator (six or seven per creator), was drawn from the analytic corpus. Two annotators with doctoral-level training in psychology coded every video independently through a purpose-built interface, blind to all model output and to each other, with presentation order shuffled. One is a female licensed psychologist specializing in children and schools; the other is a female retired psychology professor; both use various social media platforms. For each video, the interface presented a two-part coding form, with each part corresponding to one type of model output. Part A (themes) asked the annotator to tick every thematic category present, with an explicit option for none. The framework does not assign categories directly: Stage 3 produces free-form labels, and Stage 4 maps them onto the scheme by keyword, so Part A tests the open pass and the taxonomy together. Part B (indicators) asked the annotator to answer one question for each of the seven indicators in Table~\ref{tab:indicators}, choosing from a fixed set of answer options; the model answered these same questions directly and with the same options, so no mapping step is involved. The annotators and the model received different instructions. The annotators were given definitions, worked examples, and a decision rule for each item. The model was given concise question prompts, to limit input cost across the full corpus. An annotator who marked no thematic category is treated as a judgment that the video fits none of the themes, and is retained; a record with no selection at all is treated as uncoded and excluded. The first annotator returned records for all 100 sampled videos and the second for 96. Of these, three and seven, respectively, were left uncoded because the video would not play or the annotator could not follow the language, leaving 97 and 89 usable records.

\textbf{Modality ablation.} To test whether the multimodal enrichment of Stage 2 carries the agreement reported below, we coded the same 100 validation videos three further times, rebuilding each document once without the frame descriptions, once without the transcript, and once with the metadata header alone. The unified document is assembled from labeled sections, so each version deletes one section and changes nothing else; the prompt, schema, and generation settings are those of the reported run. Free-form labels from every version were mapped through the frozen keyword scheme of Stage 4, so the thematic comparison isolates the effect of modality rather than of the mapping. One metadata-only document returned no completion across three attempts and was excluded, leaving 99 videos coded in all four versions.
\section{Results}
\label{sec:results}
\subsection{Framework output and validation}\label{sec:results-framework}

\textbf{Classification output and taxonomy.} The open pass of Stage 3 produced 151,773 label instances, a mean of 15.4 per video across the 9,852 collected videos, drawn from a vocabulary of 104,026 distinct strings. On the closed pass, the model answered Unclear on 5\% or fewer of videos for five of the six binary indicators, so a single video was almost always enough to decide them. The exception is strategic hashtag use at 57.0\%: a creator's intent in choosing hashtags is rarely legible from one post. The refinement loop in Stage 4 converged on seven master categories that cover the substantive thematic domains of the study and together constitute the relevance set (Table~\ref{tab:categories}). The seven categories list 495 stems in total, 491 of them distinct, since four stems appear in two categories each. Twenty-one stems are restricted to whole-word matching, which prevents a short stem such as \emph{war} from matching inside an unrelated word such as \emph{award}. The definitions cover 27,316 distinct labels, 26.3\% of the vocabulary and 24.1\% of all label occurrences. The remainder consists largely of style and format descriptions that the closed indicators measure directly, such as \emph{Direct-to-camera monologue} and \emph{Casual conversational tone}.

The relevance restriction retains 7,311 of the 9,852 collected videos, or 74.2\%. Retention varies widely across creators, from 23.3\% for montanatucker and 28.9\% for motiankari, whose archives are dominated by general entertainment, to full retention for rawanosman2024 and hen.mazzig (Table~\ref{tab:collection}). This variation is the intended behavior of the restriction rather than a penalty on any creator, since the framework collects complete account archives while the analysis targets the advocacy output within them. Because retention is determined solely by the open labels, while the closed questions are answered independently for every video, the excluded videos provide a separate check on the restriction. The antisemitism-tied current-events indicator q3 is answered Yes for 4.1\% of the 2,541 excluded videos, compared with 57.3\% of retained videos. Manual inspection confirms that the excluded videos are predominantly dance, fashion, lifestyle, and general comedy. All statistics below are computed on the analytic corpus of 7,311 videos.

\begin{table}[!t]
  \scriptsize
  \setlength{\tabcolsep}{3pt}
  \caption{The Seven Thematic Categories Constituting the Relevance Set}
  \label{tab:categories}
  \centering
  \begin{tabularx}{\columnwidth}{@{}>{\raggedright\arraybackslash}p{0.23\columnwidth} X@{}}
    \toprule
    \textbf{Category} & \textbf{Definition capsule, with example stems} \\
    \midrule
    Geopolitics / Current Events / Media
      & Geopolitical context and international relations, current events, media coverage and bias, political analysis, international legal proceedings, campus politics and protest movements, and public debate on the Israeli-Palestinian conflict. \emph{geopolit, media bias, propaganda, ceasefire, BDS} \\
    \addlinespace
    Israel Advocacy / Pro-Israel
      & Explicit advocacy for Israel, defense of its legitimacy, Zionism, countering delegitimization, and Israeli achievement and innovation. \emph{israel, idf, zion, hasbara, jerusalem} \\
    \addlinespace
    Jewish Identity / Culture / Pride
      & Jewish identity, religious practice, holidays, language, food, communal pride, diasporic identity, and denominational diversity. \emph{jewish, torah, shabbat, kosher, hebrew} \\
    \addlinespace
    October 7 / Hamas / Hostages
      & The October 7 attack, the hostage crisis, survivor testimony, memorial content, and related security dimensions. \emph{october 7, hamas, hostage, kidnap, massacre} \\
    \addlinespace
    Counter-Antisemitism / Combating Hate
      & Identifying, confronting, and educating about antisemitism, including stereotypes and tropes, hate incidents, and harassment. \emph{antisemit, trope, blood libel, swastik, harass} \\
    \addlinespace
    Holocaust / Historical Memory
      & The Holocaust, historical persecution, collective memory, intergenerational trauma, and commemoration. \emph{holocaust, shoah, auschwitz, survivor, remembrance} \\
    \addlinespace
    Interfaith / Coexistence / Bridge-Building
      & Dialogue, coexistence, allyship, and peace-building between Jews and other religious, ethnic, or cultural groups. \emph{interfaith, coexist, allyship, dialogue, tolerance} \\
    \bottomrule
  \end{tabularx}
  \par\vspace{2pt}{\scriptsize Example stems are set in italic at the end of each capsule.\par}
\end{table}

\textbf{Agreement with expert annotators.} Table~\ref{tab:agreement} summarizes agreement between the framework and each annotator, with annotator-to-annotator agreement in the final column; Table~\ref{tab:peritem} breaks it down by item. On the closed indicators, the framework's agreement with the first and second annotator (mean $\kappa$ of 0.37 and 0.43) brackets their agreement with each other (0.42). Item by item, the framework's higher $\kappa$ exceeds the annotators' on four of the six indicators, and nearly matches it on humor (0.70 compared against  0.71). The margin is widest on call to action (0.34 against 0.16), followed by educational content (0.32 against 0.18) and response to current events (0.53 against 0.39); the first two are the items on which the annotators  agree least. These margins should be read with care: on educational content and call to action the annotators themselves agree only weakly, and the framework’s $\kappa$ of 0.32 and 0.34 remains in the fair range. Exceeding a low human benchmark indicates that the framework is no less consistent than a second expert, not that these constructs are measured reliably. Among the thematic categories, agreement is highest for October 7 / Hamas / Hostages, with $\kappa$ of 0.75 and 0.70 and raw agreement of 91.8\% and 89.9\% for the first and second annotator, respectively.

The framework's errors are systematic rather than random: relative to the annotators, it consistently overestimates the presence and intensity of what it codes. On thematic categories, the annotators reach raw agreement of 87.8\% with each other, against 75.1\% and 74.0\% between the framework and the first and second annotator. Every disagreement on whether a video carries any theme at all arises because the framework assigned one where the annotator ticked none: 18 of 97 videos for the first  annotator and 36 of 89 for the second. The framework also answers Yes far more often, for example to educational content on 61\% of the validation sample, against 9\% and 22\% for the two annotators. On emotional intensity, its score is within one point of the annotators' on 83.5\% and 69.7\% of videos on the 0 to 3 scale, but its sample mean of 1.72 is roughly double the annotators' 0.85 and 0.69. Agreement on strategic hashtag use reaches $\kappa$ = 0.55 and 0.65 but falls to 0.16 and 0.39 when restricted to videos on which both the annotator and the framework answered Yes or No rather than Unclear, suggesting that much of the agreement lies in the noncommittal answer. Despite the upward bias, average intensity per creator largely preserves the annotators' ordering: Spearman's rank correlation between the framework's creator-level means and the annotators' is 0.67 and 0.50, compared with 0.55 and 0.34 for individual videos.

\begin{table}[!t]
  \scriptsize
  \setlength{\tabcolsep}{2pt}
  \caption{Agreement Between the Framework and Two Expert Annotators}
  \label{tab:agreement}
  \centering
  \begin{tabular}{@{}llll@{}}
    \toprule
    \textbf{Measure} & \textbf{Model vs A1} & \textbf{Model vs A2} & \textbf{A1 vs A2} \\
    \midrule
    Themes, pooled over the seven categories & 75.1\% (0.37) & 74.0\% (0.32) & 87.8\% (0.50) \\
    Relevance decision confirmed             & 79 of 97  & 53 of 89  & 78.2\% (0.51) \\
    Closed indicators, mean agreement        & 73.4\%    & 76.0\%    & 85.0\% \\
    Closed indicators, mean $\kappa$         & 0.37      & 0.43      & 0.42 \\
    Emotional intensity within one point     & 83.5\%    & 69.7\%    & 95.4\% \\
    \bottomrule
  \end{tabular}
  \par\vspace{2pt}{\scriptsize Over the 97 and 89 videos each annotator coded, and the 87 both coded; three Part B items rest on 86 for the annotator pair, where one record was left blank. Percent agreement is raw, with Cohen's $\kappa$ on the binary coding used in the analysis in parentheses. A1 and A2 denote the two annotators.\par}
\end{table}

\begin{table}[!t]
  \scriptsize
  \setlength{\tabcolsep}{2pt}
  \caption{Per-Item Agreement}
  \label{tab:peritem}
  \centering
  \begin{tabular}{@{}llll@{}}
    \toprule
    \textbf{Item} & \textbf{Model vs A1} & \textbf{Model vs A2} & \textbf{A1 vs A2} \\
    \midrule
    \multicolumn{4}{@{}l}{\emph{Thematic categories, ordered by agreement with A1}} \\
    October 7 / Hamas / Hostages               & 91.8\% (0.75) & 89.9\% (0.70) & 96.6\% (0.88) \\
    Counter-Antisemitism / Combating Hate      & 76.3\% (0.40) & 74.2\% (0.28) & 89.7\% (0.52) \\
    Israel Advocacy / Pro-Israel               & 69.1\% (0.37) & 66.3\% (0.31) & 86.2\% (0.49) \\
    Geopolitics / Current Events / Media       & 63.9\% (0.27) & 62.9\% (0.22) & 88.5\% (0.44) \\
    Interfaith / Coexistence / Bridge-Building & 82.5\% (0.22) & 82.0\% (0.32) & 87.4\% (0.45) \\
    Jewish Identity / Culture / Pride          & 56.7\% (0.15) & 58.4\% (0.22) & 69.0\% (0.26) \\
    Holocaust / Historical Memory              & 85.6\% (0.11) & 84.3\% (0.18) & 97.7\% (0.49) \\
    \addlinespace
    \multicolumn{4}{@{}l}{\emph{Closed indicators, ordered by identifier}} \\
    q1 Personal storytelling                   & 83.5\% (0.40) & 79.5\% (0.26) & 90.7\% (0.38) \\
    q2 Educational content                     & 48.5\% (0.12) & 62.5\% (0.32) & 77.9\% (0.18) \\
    q3 Response to current events              & 77.3\% (0.53) & 67.0\% (0.30) & 77.9\% (0.39) \\
    q4 Call to action                          & 70.1\% (0.05) & 77.5\% (0.34) & 89.7\% (0.16) \\
    q5 Humor or sarcasm                        & 81.4\% (0.58) & 86.5\% (0.70) & 88.5\% (0.71) \\
    q6 Strategic hashtag use\textsuperscript{\dag} & 79.4\% (0.55) & 83.1\% (0.65) & 85.1\% (0.67) \\
    \bottomrule
  \end{tabular}
  \par\vspace{2pt}{\scriptsize Each row reports binary presence, for a category or for an indicator. Values are percent agreement with Cohen's $\kappa$ in parentheses. Percent agreement runs high on the low-base-rate categories, where most agreement is agreement on absence; $\kappa$ is the chance-corrected figure and the one the text argues from. \textsuperscript{\dag}q6 is inflated by the aggregation rule; see Section~\ref{sec:results}.\par}
\end{table}

\begin{table}[!t]
  \scriptsize
  \setlength{\tabcolsep}{2pt}
  \caption{Modality Ablation}
  \label{tab:ablation}
  \centering
  \begin{tabular}{@{}lcccc@{}}
    \toprule
    \textbf{Indicator} & \textbf{Full} & \textbf{No frames} & \textbf{No transcript} & \textbf{Metadata} \\
    \midrule
    q1 Personal storytelling      & 0.33 & 0.35 & 0.25 & 0.19 \\
    q2 Educational content        & 0.23 & 0.26 & 0.15 & 0.28 \\
    q3 Response to current events & 0.43 & 0.42 & 0.41 & 0.30 \\
    q4 Call to action             & 0.19 & 0.22 & 0.15 & 0.21 \\
    q5 Humor or sarcasm           & 0.63 & 0.63 & 0.59 & 0.48 \\
    q6 Strategic hashtag use      & 0.62 & 0.52 & 0.61 & 0.46 \\
    \midrule
    \textbf{Mean}                 & \textbf{0.40} & \textbf{0.40} & \textbf{0.36} & \textbf{0.32} \\
    Change from full              & ---  & $-0.003$ & $-0.045$ & $-0.086$ \\
    \bottomrule
  \end{tabular}
  \par\vspace{2pt}{\scriptsize Cohen's $\kappa$ against the human coding, averaged over the two annotators, on the 99 validation videos coded in every arm. Full is the pipeline as reported in Table~\ref{tab:peritem}. No frames drops the visual descriptions $D(v)$, No transcript drops the transcript $T(v)$, and Metadata retains the header alone.\par}
\end{table}

Table~\ref{tab:ablation} reports agreement for the four versions of the modality ablation, on the 99 videos coded in all four. Mean $\kappa$ across indicators and annotators is 0.40 for the full document, 0.40 without the frame descriptions, 0.36 without the transcript, and 0.32 with the metadata header alone. The version without frames thus retains nearly all of the full document's agreement, and the version without the transcript most of it; a clear drop appears only when both are removed, a relative decline of 21\% concentrated on the items that depend on what a video says or shows. Strategic hashtag use is the clearest case: its  $\kappa$ falls from 0.62 to 0.46 even though the hashtags themselves are in the metadata header, because judging whether they are used strategically requires the content they accompany. Educational content and call to action move in the opposite direction, gaining agreement when content is removed. This is consistent with the framework's tendency to over-assign Yes on these items: with less content to draw on, it answers Yes less often, closer to the annotators' rates.  These are also the two items on which the annotators agree least, so the shift should not be over-read.

The open pass shows a trade between coverage and precision. Enrichment, that is, adding the transcript and frame descriptors to the metadata header, recovers more of the themes the annotators identified, 88 and 64 correct assignments for the full document, against 71 and 53 with the metadata header alone. It also produces more false positives, so the annotators agree with the framework on whether a video carries any theme less often for the full document, on 81\% and 60\% of videos against 95\% and 83\% on metadata alone. Because every thematic prevalence reported here is already read as an upper bound, the recall that enrichment gains matters more than the precision it costs.

\subsection{Communication patterns of the sixteen creators}\label{sec:results-creators}

\begin{figure*}[!t]
  \centering
  \includegraphics[width=\textwidth]{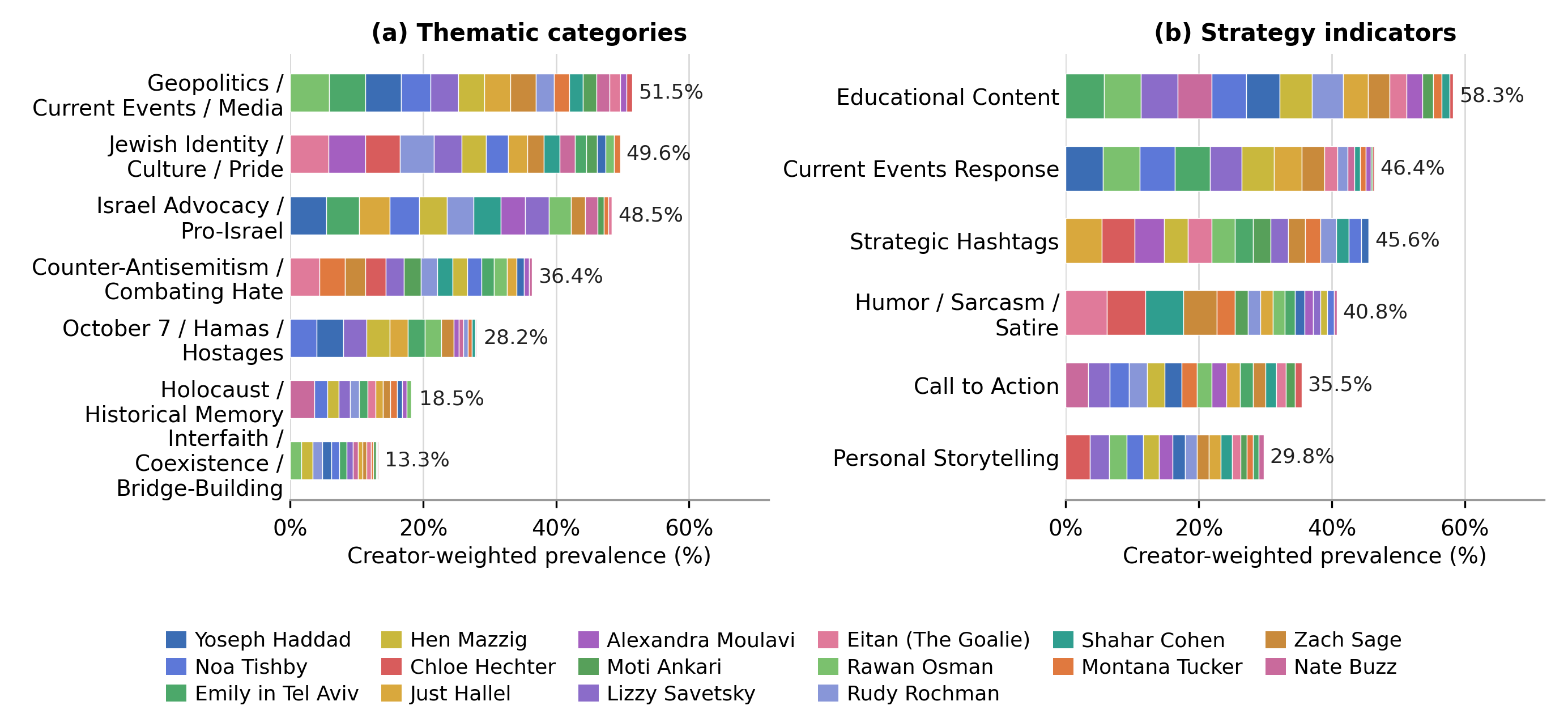}
  \caption{Creator-weighted prevalence of (a) the seven thematic categories and (b) the six binary indicators across the analytic corpus of 7,311 videos from sixteen creators. Each bar decomposes the creator-weighted mean into per-creator contributions, each the proportion of the creator's videos in which the category or indicator is present, divided by sixteen, so that the full bar equals the creator-weighted prevalence. Colors identify creators consistently across figures.}
  \label{fig:prevalence}
\end{figure*}

\textbf{Thematic profile.} Figure~\ref{fig:prevalence}a reports creator-weighted thematic prevalence, and Table~\ref{tab:theme-prevalence} provides the underlying values. The three leading themes, Geopolitics, Jewish Identity, and Israel Advocacy, cluster within three points of one another at around half the corpus, after which prevalence falls away steadily to 13.3\% for interfaith content. The corpus thus mixes a stable identity-centered register with an event-driven advocacy register. Different creators carry the two registers. Jewish Identity is broadly distributed, with the largest per-creator shares among eitanthegoalie at 93.0\%, alexandramoulavi at 88.7\%, and chloehechterr at 82.0\%. In contrast, the event-driven themes concentrate in the news-oriented accounts, with yosephhaddad reaching 87.2\% on Israel Advocacy, 85.5\% on Geopolitics, and 62.9\% on October 7. Holocaust / Historical Memory is led by natebuzz, who devotes 58.0\% of his advocacy output to it.

\begin{table}[!t]
  \scriptsize
  \caption{Prevalence of the Seven Thematic Categories}
  \label{tab:theme-prevalence}
  \centering
  \begin{tabular}{@{}lrrr@{}}
    \toprule
    \textbf{Category} & \textbf{CW} & \textbf{SD} & \textbf{VW} \\
    \midrule
    Geopolitics / Current Events / Media       & 51.5\% & 25.8 & 58.7\% \\
    Jewish Identity / Culture / Pride          & 49.6\% & 26.2 & 45.9\% \\
    Israel Advocacy / Pro-Israel               & 48.5\% & 27.7 & 56.7\% \\
    Counter-Antisemitism / Combating Hate      & 36.4\% & 16.8 & 33.9\% \\
    October 7 / Hamas / Hostages               & 28.2\% & 23.6 & 37.8\% \\
    Holocaust / Historical Memory              & 18.5\% & 14.0 & 14.9\% \\
    Interfaith / Coexistence / Bridge-Building & 13.3\% & 8.7  & 14.1\% \\
    \bottomrule
  \end{tabular}
  \par\vspace{2pt}{\scriptsize CW is the creator-weighted mean, SD its standard deviation across the sixteen creators, and VW the video-weighted pooled rate over the analytic corpus.\par}
\end{table}

    \textbf{Advocacy strategies.} Table~\ref{tab:indicators} reports the prevalence of the seven indicators. Among the six binary indicators, educational content leads at 58.3\% (creator-weighted), followed by response to current events at 46.4\%; the remaining four range from 45.6\% for strategic hashtag use down to 29.8\% for personal storytelling. Creators differ most by response to current events and humor or sarcasm (standard deviations of 34.1 and 31.2 percentage points) and least on calls to action and personal storytelling (10.7 and 12.9). Weighting matters most for response to current events: its video-weighted rate of  57.3\% exceeds  the creator-weighted 46.4\%, indicating that creators who post more also respond to current events more. Mean emotional intensity is 1.78 on the 0 to 3 scale. Three figures warrant caution. Strategic hashtag use is answered Unclear on 57.0\% of videos, which the aggregation rule counts as absence; with a video-weighted Yes rate of 42.0\%, only about 1\% are answered No, so for this indicator the model in effect chooses between Yes and Unclear. The framework also overstates both educational content and emotional intensity relative to the annotators (Section~\ref{sec:results-framework}).
    
\begin{table}[!t]
  \footnotesize
  \caption{The Seven Reported Indicators and Their Prevalence}
  \label{tab:indicators}
  \centering
  \begin{tabular}{@{}llrrrr@{}}
    \toprule
    \textbf{ID} & \textbf{Indicator} & \textbf{Unc.} & \textbf{CW} & \textbf{SD} & \textbf{VW} \\
    \midrule
    q2 & Educational content        & 0.5\%  & 58.3\% & 28.9 & 60.1\% \\
    q3 & Response to current events & 4.9\%  & 46.4\% & 34.1 & 57.3\% \\
    q6 & Strategic hashtag use\textsuperscript{\dag} & 57.0\% & 45.6\% & 21.9 & 42.0\% \\
    q5 & Humor or sarcasm           & 1.8\%  & 40.8\% & 31.2 & 41.9\% \\
    q4 & Call to action             & 1.8\%  & 35.5\% & 10.7 & 35.4\% \\
    q1 & Personal storytelling      & 0.1\%  & 29.8\% & 12.9 & 33.6\% \\
    q7 & Emotional intensity (0 to 3) & ---  & 1.78   & 0.51 & 1.93   \\
    \bottomrule
  \end{tabular}
  \par\vspace{2pt}{\scriptsize Unc.\ is the share of analytic-corpus videos the model answered Unclear, which the aggregation rule pools with absence. CW is the creator-weighted mean, SD its standard deviation across the sixteen creators, and VW the video-weighted pooled rate over the analytic corpus. q1 through q6 use the scale Yes / No / Unclear; q7 is ordinal 0 to 3 and admits no Unclear response. \textsuperscript{\dag}q6 is answered Unclear on more than half the corpus, so its prevalence and its agreement figures both rest on that pooling; see Section~\ref{sec:results}.\par}
\end{table}

\textbf{Creator profiles.} Figure~\ref{fig:prevalence}b decomposes each prevalence into per-creator contributions and reveals two recurring archetypes. A news-driven educator cluster of six accounts pairs high educational content with high responsiveness to current events, most sharply emilyintelaviv at 91.9\% and 83.8\% respectively. A comparable entertainment-native cluster instead builds advocacy on humor and platform vernacular, led by eitanthegoalie at 99.2\% humor. The two are visible in Figure~\ref{fig:prevalence}b as different creators filling the same bar: corpus-level prevalence that looks alike is produced by opposite kinds of accounts.

\begin{figure}[!t]
  \centering
  \includegraphics[width=\columnwidth]{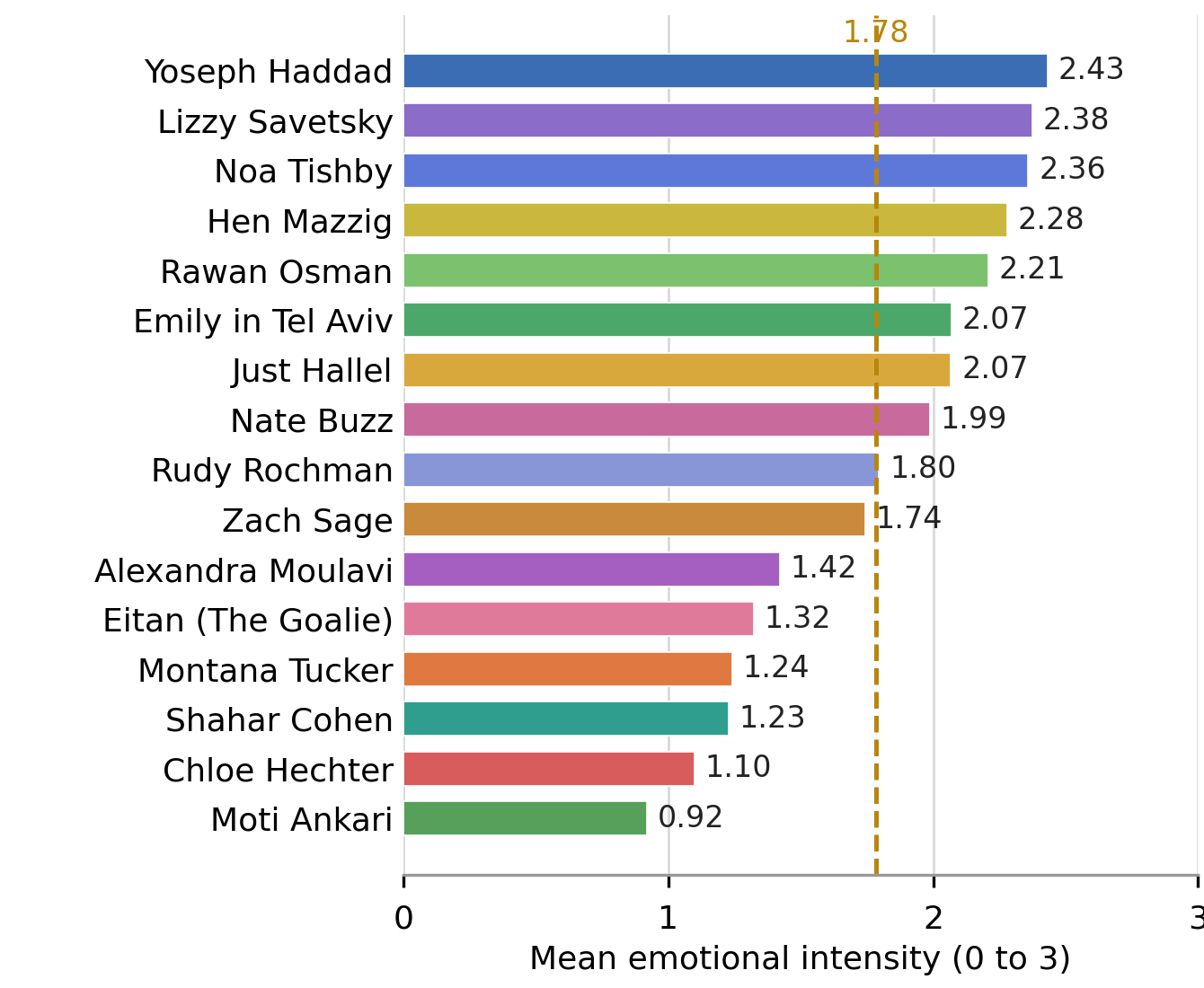}
  \caption{Mean emotional intensity (q7, ordinal scale 0 to 3) per creator over the analytic corpus. The dashed line marks the creator-weighted corpus mean of 1.78.}
  \label{fig:intensity}
\end{figure}

The creator-weighted mean intensity is 1.78 on the 0 to 3 scale, indicating that moderate emotional framing is the corpus norm. The distribution across creators follows the same divide as the strategies. The news-driven educators occupy the top of the scale, with yosephhaddad at 2.43, lizzysavetsky at 2.38, noatishby at 2.36, hen.mazzig at 2.28, and rawanosman2024 at 2.21, while the entertainment-native creators fall far below the corpus mean, with motiankari at 0.92, chloehechterr at 1.10, shaharcohen9 at 1.23, montanatucker at 1.24, and eitanthegoalie at 1.32 (Figure~\ref{fig:intensity}).

\textbf{Effect of the normalization.} The creator-weighted and video-weighted estimates diverge exactly where the volume imbalance of Table~\ref{tab:collection} predicts. The largest correction concerns the response to current events, which falls from 57.3\% video-weighted to 46.4\% creator-weighted because the highest-volume accounts are precisely the news-driven ones, and emotional intensity likewise falls from 1.93 to 1.78. In the opposite direction, hashtag use rises under creator weighting because the creators who use hashtags most heavily post at lower volume. These shifts confirm the choice of the creator-weighted mean as the primary statistic, since the pooled rate would otherwise reflect the largest accounts rather than the creator population. That choice also equalizes creators whose audiences differ by three orders of magnitude, from a median of 1,361 plays per video to 875,750 (Table~\ref{tab:collection}). The equalization is deliberate, since the framework measures what a creator does rather than how far it travels.

\section{Discussion and limitations}
\label{sec:discussion}

We developed an automated framework for collecting and analyzing short-form videos, and our results show that it reaches expert-level agreement while covering an entire corpus rather than a hand-coded sample. Its agreement with expert annotators matched their agreement with each other, and when results were aggregated per creator, the level at which the framework is meant to be used, it ranked creators much as the experts did. The framework collected close to 10,000 videos, about 155 hours of footage, from sixteen creators who differ widely in terms of demographics, style, and audience size, spanning more than seven years (Table~\ref{tab:collection}). The framework analyzed all of these videos automatically, identifying the topics each creator addresses and the communication strategies each one uses, and produced a profile for each creator that can be compared with those of the others (Tables~\ref{tab:theme-prevalence} and~\ref{tab:indicators}, Figures~\ref{fig:prevalence} and~\ref{fig:intensity}). Because the framework collects every public video from each account, not a sample, these profiles reflect how the creators actually work and allow us to learn how influencers operate based on the content they have published. Analysis at this scale was not practical before LLMs became available to researchers. We tested the framework on one platform, TikTok, but, since it relies on the video itself and its metadata, it can be applied to other video platforms, such as Instagram.

The analysis itself rests on two kinds of questions that the model answers for every video. The open questions ask the model to  describe freely the video's topics and how it is presented, and the themes are built from these descriptions.  The framework thus does not only measure themes defined in advance but also discovers new ones: some were derived from the literature, while others emerged from the data (Table~\ref{tab:categories}). The closed questions, by contrast, offer a fixed set of answers, for example, whether the creator uses humor or tells a personal story, which  allows systematic comparison across creators (Table~\ref{tab:indicators}). These questions yielded three general findings. First, the framework clearly separates videos on the seven themes from all other videos: the closed question on responses to current events, which was also answered for the excluded videos, received a Yes for 57.3\% of included videos  but for only 4.1\% of the excluded ones, which dealt mostly with dance, fashion, lifestyle, and general comedy. Second, creators differ greatly in how they deliver their message: the use of educational content and humor varies widely from creator to creator, while calls to action appear at similar rates for all of them (Figure~\ref{fig:prevalence}). Third, prevalences should be computed at the creator level. When each video is counted separately, a single prolific account can skew the overall result;  in our case, the rate of response to current events fell from 57.3\% to 46.4\% once each creator was given equal weight.

Compared with those of human annotators, the framework's results are similar. On the closed questions, the agreement between the two annotators ($\kappa = 0.42$) falls between the framework's agreement with each of them (0.37 and 0.43). On four of the six indicators, the framework agreed with at least one annotator more than the annotators agreed with each other (Tables~\ref{tab:agreement} and~\ref{tab:peritem}). Agreement is not uniform across indicators, and it is possible to identify when the framework works better and when it works less well. It works especially well when the answer can be found directly in what is said or displayed in the video, as with humor or response to current events. It works less well when the concept itself is ambiguous: agreement is low for educational content and the call to action, but the two annotators also agreed with each other on these items at $\kappa$ values of only 0.18 and 0.16. Hence, the difficulty arises from the definition, not from the model. It also works less well when the question concerns an intent that is not visible in a single video, such as the strategic use of hashtags. 

For emotional intensity, the framework assigns higher scores than the annotators do, partly because the model was asked how emotionally charged the video's content is. In contrast, the annotators were asked how emotional its delivery is and told to ignore the seriousness of the topic. It nonetheless ranks creators much as the annotators do, especially when each creator's mean is compared (Spearman's rank correlation of 0.67 and 0.50 with the two annotators). For the open questions that pass through the mapping to categories, agreement is lower (75.1\% and 74.0\%, compared with 87.8\% between the annotators), for reasons discussed below. Agreement on the weaker items could perhaps be improved with different prompts, for example, with definitions and examples like those the annotators received, and this is worth exploring in future work.

Beyond agreement itself, the framework makes possible what human coders cannot do in practice. Based on the annotators' coding times, manually coding all 155 hours of video in the corpus would take about 600 hours of work, roughly seventy-five working days for a single annotator. The framework performs this coding automatically, and the same process can be applied to thousands of hours of video, yielding insights that manual coding cannot reach.

The modality ablation shows that, in the case we studied, combining the video's modalities improves the results. The framework converts each video into a single text document that combines the metadata, the speech transcript, and the frame descriptions (Figure~\ref{fig:framework}). When the model received the full document, mean agreement with the annotators on the closed questions was highest, at $\kappa = 0.40$; when only the metadata remained, it fell to 0.32, a relative decline of 21\% (Table~\ref{tab:ablation}). On the open questions, too, the full document recovered more of the topics the annotators identified. Partial information, however, still yields good results: without the frame descriptions, agreement remained at 0.40, and without the transcript, it was 0.36. The transcript and the frame descriptions thus largely compensate for each other's absence, so the framework continues to work well even when one is missing, for example in a video without speech, and it can run in a cheaper configuration when frame description is costly. The small contribution of the frames stems in part from the kinds of questions examined, which mostly concern what is said. At the level of the labels themselves, the frames capture cues that the transcript cannot: in the validation sample, 18 videos received labels describing visual symbols or clothing in the full version, such as flags or writing on clothes, and without the frames, 11 of them lost these labels. For visual or more detailed questions, the frame descriptions are therefore likely to matter more.

The results could be improved further in several ways. One is to use a stronger vision--language model for the frame descriptions (this study used a relatively small model, Qwen2.5-VL 7B), or a multimodal model that takes the video itself, images, and audio, rather than a text document that summarizes it. A second is to add audio cues that the transcript does not capture, such as tone of voice and music, which may be particularly helpful for questions about emotional intensity. A third is to combine several language models and decide by their agreement, reducing the errors of any single model. Future work should also compare open and closed models at each stage of the framework. The enrichment settings themselves are worth testing as well. Our experiment sampled one frame per second, which greatly reduces the risk of missing visual information but makes the unified document too long for long videos. As a result, documents exceeding 20,000 characters had their frame descriptions evenly subsampled to at most 30 frames. For long videos, the model therefore saw only part of the frames, and using more frames may improve the results further, a direction worth testing in future work. A better way of choosing which frames enter the document, for example, frames at scene changes or frames that add new information over the preceding ones, could preserve the important information while keeping the document short.

The framework also has limitations, and for the most part, it is possible to anticipate when it will work better and when it will work less well. It works well when the answer to a question can be found in what is said or shown in the video itself. It works less well when the question concerns an intent that becomes clear only across many posts, such as the strategic use of hashtags, or a concept so ambiguous that people also struggle to agree on it. It also struggles with behaviors that are not visible in a published video, such as live streams or posting times, and with questions whose answer depends on tone and delivery, which the model does not hear. The weakest component of the framework is the keyword mapping from free-text labels to categories. It was chosen for its transparency, since every assignment can be traced to the keyword that produced it, but it over-assigns through two main mechanisms. The first is a keyword that appears inside another word: \emph{petition} occurs inside \emph{repetition}, so the label ``Repetition for emphasis'' was assigned to the Geopolitics category. The second is that the mapping ignores negation: a label stating that a topic does not appear in the video is assigned to that very topic. Topic prevalences should therefore be read as upper bounds.

Most of these limitations can be addressed. The mapping can be improved by filtering out negated labels and matching whole words only, or replaced by semantic matching, for example with text embeddings.  A more promising direction is to drop the mapping altogether. Since the framework agrees with the annotators on the closed questions about as well as they agree with each other, the open questions can be run to discover the themes, and each theme can then be posed as a closed question for every video. This would also need to be checked for the upward bias the framework shows on some closed questions.  

Applied to the sixteen influencers, the framework yields one central finding: two creator types recur across all measures. In topics, strategies, and emotional intensity, the same creators fall into the same group repeatedly (Figure~\ref{fig:prevalence}). The first type is the news commentator: six accounts that combine a high rate of educational content with responsiveness to current events, led by emilyintelaviv, with 91.9\% educational content and 83.8\% response to current events. Current-events topics are concentrated in these accounts: for yosephhaddad, for example, 87.2\% of videos address support for Israel and 85.5\% address geopolitics. These accounts also stand out for their emotional intensity: the five creators with the highest intensity, ranging from 2.43 for yosephhaddad to 2.21 for rawanosman2024, all belong to this type (Figure~\ref{fig:intensity}). The second type is the entertainment creator, who delivers the message through humor and in the platform's own formats, exemplified by eitanthegoalie, with 99.2\% of videos using humor or sarcasm. Two of these creators, eitanthegoalie and chloehechterr, also have some of the highest rates of Jewish identity content, at 93.0\% and 82.0\%, and the five creators with the lowest intensity, ranging from 0.92 for motiankari to 1.32 for eitanthegoalie, all belong to this type. The two types demonstrate that there is no single way to deliver a public message on the platform, and a mean across all creators hides this division: the standard deviation across creators reaches 28.9 percentage points for educational content and 31.2 for humor, against only 10.7 for calls to action, which appear at comparable rates for every creator.

The framework’s accuracy was not uniform across topics (Table~\ref{tab:peritem}). It was strongest on the October 7 / Hamas / Hostages category, a focused topic with a distinctive vocabulary, where agreement with the two annotators reached $\kappa$ values of 0.75 and 0.70. It performed less well on broad, diffuse topics, such as Jewish identity, on which the annotators themselves agreed only weakly ($\kappa = 0.26$), and in categories affected by mapping errors, such as Geopolitics. The division into two types is nonetheless robust because it recurs across several separately measured indicators. These include humor, on which the framework reached its highest agreement among the closed questions, and emotional intensity, on which its ranking of creators broadly matches the annotators’ (Spearman’s $\rho$ = 0.67 and 0.50). 

Beyond the two types, the analysis produced several additional findings about the influencers. First, the share of a creator's videos that address the seven themes (the retention rate in Table~\ref{tab:collection})  is related to how these themes are delivered. In the montanatucker account, most of whose content is general entertainment, only 23.3\% of videos address these themes, whereas in the rawanosman2024 and hen.mazzig accounts, all videos do (Table~\ref{tab:collection}). The Spearman correlation between this share and the mean emotional intensity of the videos is 0.74: a creator for whom these themes are only part of their world presents them in a moderate tone. In contrast, a creator for whom they are the main focus presents them with greater intensity. The pattern holds, with a correlation of 0.66, even after removing the videos included in the analysis only because of the mapping errors. Second, the most frequent labels describe form and not topic, led by direct address to the camera in a conversational tone, while platform-specific tools such as duets, stitches, green screen, and live streams are hardly used to deliver the message. The creators thus rely on themselves more than on production techniques. Third, audience size does not separate the two types: the accounts differ greatly in reach, from 1,361 to 875,750 views for the median video, and both types include small and large accounts.

The framework also enables us to address questions that are difficult to answer with manual coding, and two examples from our data illustrate this. The first is tracking change over time. Because every video carries a publication date, periods can be compared, revealing a sharp change after October 7, 2023: before that date, 55.9\% of all published videos addressed at least one of the seven themes, and afterward, 89.9\%. Of the 15 creators who posted both before and after that date, 13 increased the share of their videos addressing these themes, and the remaining two had already focused almost exclusively on them. The change is especially pronounced among entertainment creators: for chloehechterr, for example, the share of the account's videos addressing these themes rose from about a third to more than two-thirds. Emotional intensity also rose after October 7 for each of these fifteen creators, and the rawanosman2024 archive begins only in October 2023, with every video addressing these themes. The second example is the relationship between strategy and views. Comparing videos by the same creator, humorous videos received more views than other videos for 13 of 14 creators, typically about a third more, while no other strategy showed a similar advantage. Such findings do not establish causality, but they show that the framework enables examination not only of what influencers do but also of what works for them, and makes it possible to extend these questions to other populations and platforms.

The specific analysis of the influencers also has its own limitations. First, it describes sixteen accounts in a single domain and on a single platform, drawn from a list we compiled rather than sampled at random, so its findings cannot be generalized to all creators working on these topics. The corpus also has properties that shape the results: posting volume differs greatly across accounts, one account contributes about a third of the analyzed videos, and most of the analyzed videos were published after October 7, 2023. Second, the findings describe what the creators do, but do not explain why. We cannot determine whether the two types are stable over time or a response to the period after October 7, and the relationships we found, for example, between humor and views, do not establish causality. Third, some of the measurements in this analysis call for caution. Topic prevalences are upper bounds because of the mapping's over-assignment. The overall mean of emotional intensity is higher than that of the annotators. Strategic hashtag use is not among the validated indicators, and the eleven sparse items, including platform-specific tools, were not validated by the annotators. In addition, the validation set of 100 videos is too small to assess the framework's accuracy for each creator individually. Most of these limitations can be reduced in future work: repeated snapshots of the same accounts would test stability over time, running the framework on other populations and platforms would test whether the patterns generalize, and a larger validation sample would allow the framework to be tested at the level of the individual creator.

Even within these limits, however, the analysis reveals insights about the influencers studied that would be very challenging to reach through manual coding: two clear creator types that recur in topics, strategies, and emotional intensity, and a sharp change in their behavior after October 7, 2023. The findings are also robust: the relationship between the share of a creator's output devoted to the seven themes and emotional intensity holds even after removing the videos included in the analysis by mistake. Together, these findings illustrate the value of analyzing complete archives over years and the framework's potential for studying influencers well beyond the case examined here.

\section{Conclusions and future work}
\label{sec:conclusions}

In this work, we presented a novel framework for the automated analysis of creator communication on short-form video platforms. The framework collects each creator's complete archive and combines what is said in each video, what appears on screen, and its metadata into a single text document. It then analyzes that document with a language model through two kinds of questions: open questions, which discover the topics creators address, and closed questions, which measure their communication strategies consistently. From the answers, it produces a profile for each creator that can be compared with those of other creators.

We tested the framework on close to 10,000 videos from sixteen TikTok creators. For the closed questions, it reached the agreement level of trained human annotators at a scale and speed that manual coding cannot match, and the modality ablation showed that combining the video's modalities improves results. The analysis revealed two types of influencers, news commentators and entertainment creators, and a sharp change in the content and emotional intensity of their videos after October 7, 2023.

What makes the framework distinctive is the combination of three principles: it operates on complete archives, not samples; it merges several modalities (text, speech, and image) into a single document; and it combines open and closed questions. Because it is not tied to a particular domain, it transforms the study of influencer communication from a manual, limited task into a quantitative, broad, and ongoing field of research.

There are several directions for future research. The framework can already be applied to other creator populations and to other video platforms, such as Instagram. It can also be improved by using more frames and other models, open and closed, by adding audio signals, and by replacing the keyword mapping with closed questions about each category. Repeated snapshots and a larger validation sample would extend the analysis over time and to the level of the individual creator. The code for the analysis stages will be publicly available on GitHub upon publication, so researchers can adapt the framework to their own questions and domains and replace any of its models.

As more and more public discourse moves to short-form video, the ability to measure it systematically, across complete archives and over time, becomes essential. The framework presented here shows that this is already possible and opens a way to understand not only what influencers say but also how they say it and how this changes over time.

\section*{Code Availability}
The implementation will be made publicly available upon publication at \url{https://github.com/omertarshish/creator-measurement-pipeline}. It covers Stages 2 to 5, includes the prompts, indicator definitions, and keyword-scheme loader, and is configured so that the speech, vision--language, and text models can each be replaced. Links to all analyzed videos will also be made available upon publication. The video corpus is not redistributed: it is third-party content subject to platform terms, and creator archives change as videos are edited or withdrawn.

\section*{Acknowledgements}
This work was supported by an NJIT-BGU research grant. We thank the two annotators for their careful coding of the validation sample.

\bibliographystyle{unsrt}
\bibliography{references}

\sheading{Appendices}

\appheading{Vision--language frame-description prompt (Stage 2)}{app:vlprompt}

Each sampled frame is passed to the vision--language model together with the instruction reproduced verbatim below, which elicits a structured natural-language description rather than a default caption.

\begin{promptblock}
Analyze this TikTok video frame with comprehensive detail for marketing and content insights. Provide thorough analysis in these sections: (1)~PEOPLE \& INFLUENCERS: Describe all people - appearance (age, gender, hair, clothing, makeup), expressions, body language, actions, and interactions. Note influencer characteristics and audience appeal factors. (2)~CONTENT \& THEME: Identify content type (dance, comedy, educational, lifestyle, etc.), mood, narrative context, cultural elements, and target audience indicators. (3)~VISUAL ELEMENTS: Describe setting, lighting, color palette, camera work, composition, visual effects, filters, and production quality. (4)~TEXT \& BRANDING: Transcribe ALL visible text exactly. Include captions, hashtags, usernames, brand names, product names, and any overlay text. For non-English text, provide original and translation. (5)~MARKETING ANALYSIS: Identify brand elements, product placements, sponsored content indicators, call-to-actions, engagement tactics, and viral potential factors. (6)~CULTURAL CONTEXT: Note trends, challenges, memes, cultural references, music/audio cues, and social context. Be extremely detailed, accurate, and complete. Focus on elements relevant for marketing analysis and trend identification.
\end{promptblock}

Each request prefixes this instruction with a positional header of the form [TikTok Frame i/N at t s], giving the model the index of the frame within the video and its timestamp. From the second frame of a video onward, a consistency clause is appended after the instruction, listing person, brand, and setting features recovered from earlier frames of the same video in the form ``For consistency with previous frames: \ldots{} Please maintain consistent descriptions.'' Generation used temperature 0.05, top-p 0.92, top-k 40, repetition penalty 1.03, and a 700-token output limit; frames were resized with cubic interpolation.

\appheading{Classification prompts and response schema (Stage 3)}{app:prompts}

The two passes of Stage 3 present the same unified video document $U(v)$ to the model. The instruction text of both passes is reproduced below, and in each case the user message contains $U(v)$.

\smallskip\noindent\emph{Open-question pass.}

\begin{promptblock}
You are analyzing a TikTok video produced by a creator active in Jewish advocacy, Israeli public diplomacy, or counter-antisemitism content.

LABELS

Produce a single list that captures: (1) the key topics, subjects, and ideas the creator addresses, and (2) the communication strategies and rhetorical approaches they use to convey their message. Each item must be a concise label of no more than five words. Each label must be distinct. Do not include labels that overlap or convey the same idea in different words.
\end{promptblock}

\smallskip\noindent\emph{Closed-question pass.}

\begin{promptblock}
You are analyzing a TikTok video produced by a creator active in Jewish advocacy, Israeli public diplomacy, or counter-antisemitism content.

CLOSE QUESTIONS

Answer each of the following questions about the video. Use ONLY the allowed answers shown in parentheses. Items 1 to 6 admit Yes / No / Unclear. q1~\emph{personal\_storytelling}: Does the creator use personal storytelling or narratives? q2~\emph{educational}: Does the creator create educational videos? q3~\emph{current\_events}: Does the creator respond to current events or viral content related to antisemitism? q4~\emph{call\_to\_action}: Does the creator use calls to action? q5~\emph{humor\_sarcasm}: Does the creator use humor, sarcasm, or satire? q6~\emph{strategic\_hashtags}: Does the creator use hashtags strategically around antisemitism awareness? q7~\emph{emotional\_intensity}: Rate the emotional intensity of this video's advocacy content (0 = purely factual/neutral, 1 = mild emotional coloring, 2 = moderate emotional framing, 3 = emotion is the primary persuasion strategy)
\end{promptblock}

\smallskip\noindent\emph{Closed questions not reported here.} The reported run asked eleven further closed questions in the same instruction. They are not reported in the tables because too few videos were answered Yes (Section~\ref{sec:experiments}), and all admit Yes / No / Unclear unless noted.

\begin{promptblock}
\emph{trending\_formats}: Does the creator use trending sounds, formats, or challenges to embed messages about antisemitism? \emph{series\_format}: Does the creator produce content in a series format? \emph{data\_statistics}: Does the creator use data or statistics? \emph{green\_screen}: Does the creator use the green screen effect to visually reference news articles, images, or other sources about antisemitism? \emph{longer\_form\_redirect}: Does the creator direct followers to longer-form content on other platforms about antisemitism? \emph{live\_conversations}: Does the creator go live on TikTok to have real-time conversations with followers about antisemitism? \emph{strategic\_timing}: Does the creator post content at strategic times to maximize algorithmic reach for their content? \emph{tagging\_accounts}: Does the creator tag and mention other accounts? \emph{duets\_stitches}: Does the creator use duets or stitches to respond to or amplify other creators' content related to antisemitism? \emph{collaborations}: Does the creator collaborate with other creators about antisemitism awareness? \emph{trend\_for\_advocacy}: Does this video adopt a widely-recognized social media content format, template, or participatory trend to deliver a message about antisemitism, Jewish identity, or Israel advocacy? (Yes / No / Uses a trend format but not for advocacy / Unclear)
\end{promptblock}

Responses are constrained by a strict JSON schema in structured-output mode with strict enforcement. The open pass returns a single object with a labels field holding an array of strings. The closed pass returns a single object with one field per question, where q1 through q6 are restricted to the enumeration \{Yes, No, Unclear\} and q7 to the integers \{0, 1, 2, 3\}. All fields are required and additional properties are prohibited, so every video yields a syntactically valid and complete record. Generation parameters were temperature 0.0, top-p 1.0, and a 4,096-token output limit.

\appheading{Taxonomy construction (Stage 4)}{app:taxonomy}

This appendix details the label-to-category mapping summarized in Stage 4.

\begin{pseudo}
$\mathcal{L} = \cup_v L(v)$ \quad the corpus vocabulary, the distinct labels produced across all videos\par
$\mathrm{freq}(l) = |\{v : l \in L(v)\}|$ \quad the number of videos in which label $l$ occurs\par
$S_m$ \quad the keyword stems of category $C_m$, base word forms realizing its natural-language criterion\par
$\mathrm{match}(s, l) = 1$ \quad when stem $s$ occurs in label $l$, matched case-insensitively, and 0 otherwise\par
$\mathrm{assigned}(l, C_m) = 1$ \quad when some $s \in S_m$ satisfies $\mathrm{match}(s, l) = 1$
\end{pseudo}

Because the stems are truncated base forms, substring containment lets a single stem capture its morphological variants. Word-boundary anchoring is not the default and is introduced only selectively, in the refinement loop below, to constrain individual broad stems that would otherwise match unrelated words.

The stem sets are not fixed in advance but are developed through the following auditable refinement loop.

\begin{pseudo}
\textbf{Input:} vocabulary $\mathcal{L}$ with video frequencies $\mathrm{freq}(\cdot)$\par
\pin{3.6} provisional categories with initial stem sets $S_m$\par
\textbf{Output:} frozen stem sets $S_m$ \par\vspace{0.35\baselineskip}
\textbf{repeat}\par
\pin{2} for each category $C_m$:\par
\pin{4} assign every label $l$ in $\mathcal{L}$ for which some stem in $S_m$ occurs in $l$,\par
\pin{6} recording the triggering stem for each assignment\par
\pin{4} inspect assignments, ordered by $\mathrm{freq}(l)$ and grouped by triggering stem\par
\pin{4} for each systematic miss (label belongs to $C_m$ but was not assigned):\par
\pin{6} add a stem to $S_m$ that captures it\par
\pin{4} for each spurious match (label assigned through incidental overlap):\par
\pin{6} replace the offending stem with a word-boundary-anchored form\par
\textbf{until} a full pass yields no material change in the assignments\par
freeze the stem sets $S_m$ and the category criteria
\end{pseudo}

Two error types are corrected at each pass. Systematic misses, in which labels that clearly belong to a category are not captured, are remedied by adding stems. Spurious matches, in which a broad stem captures unrelated labels through incidental substring overlap, are remedied by replacing the offending stem with a word-boundary-anchored pattern, so that, for instance, a stem intended to capture references to art is constrained to match the word in isolation rather than as a fragment of unrelated terms such as ``start,'' ``party,'' or ``smart.'' The loop terminates when a further pass produces no material change in the assignments, at which point the category definitions and their keyword sets are frozen and applied unchanged to every video in the corpus. Because the frozen definitions and the per-label triggering stems are retained, the mapping is deterministic and reproducible given the same inputs, rather than the product of undocumented coder judgment. The frozen stem sets are available from the authors on request.

\appheading{Normalization and aggregation (Stage 5)}{app:normalization}

\begin{pseudo}
\textbf{Input:} videos $v = 1, \ldots, N$, each with creator $k(v)$ and mapped category set $C(v)$\par
\pin{3.6} creators $k = 1, \ldots, K$ with video sets $V_k$\par
\pin{3.6} categories $c = 1, \ldots, M$ \par\vspace{0.35\baselineskip}
\textbf{Step 1} (binary presence, removes label redundancy):\par
\pin{2} for each video $v$ and category $c$:\par
\pin{4} $I(v, c) \leftarrow 1$ if $c \in C(v)$, else 0\par
\textbf{Step 2} (creator-level proportions):\par
\pin{2} for each creator $k$ and category $c$:\par
\pin{4} $p(k, c) \leftarrow \frac{1}{|V_k|} \sum_{v \in V_k} I(v, c)$\par
\textbf{Step 3} (creator-weighted aggregation, primary statistic):\par
\pin{2} for each category $c$:\par
\pin{4} $p_{\mathrm{CW}}(c) \leftarrow \frac{1}{K} \sum_{k=1}^{K} p(k, c)$ \par\vspace{0.35\baselineskip}
report $p_{\mathrm{CW}}(c)$ together with the SD of $p(k, c)$ across creators
\end{pseudo}

\end{document}